\documentclass[12pt]{article}

\usepackage{amsmath,amssymb,amsfonts}
\usepackage{url}
\usepackage{graphicx}
\usepackage{tikz}
\usepackage{cite}
\usepackage{multicol}
\usepackage[normalem]{ulem}
\usepackage[pagewise]{lineno} 

\begin{document}
\title{Mathematical model of SARS-Cov-2 propagation versus ACE2 fits COVID-19 lethality across age and sex and predicts that of SARS, supporting possible therapy}
\author{Ugo Bastolla$^{(1)}$\\
  \small $^{(1)}$ Centro de Biologia Molecular "Severo Ochoa"\\
  \small CSIC-UAM Cantoblanco, 28049 Madrid, Spain. 
\small email: ubastolla@cbm.csic.es
}

\date{}
\maketitle

\begin{abstract}
The fatality rate of Covid-19 escalates with age and is larger in men than women. I show that these variations correlate strongly with the level of the viral receptor protein ACE2 in rat lungs, which is consistent with the still limited and apparently contradictory data on human ACE2. Surprisingly, lower levels of the receptor correlate with higher fatality. However, a previous mathematical model predicts that the speed of viral progression in the organism has a maximum and then declines with the receptor level. Moreover, many manifestations of severe CoViD-19, such as severe lung injury, exacerbated inflammatory response and thrombotic problems may derive from increased Angiotensin II (Ang-II) level that results from degradation of ACE2 by the virus.
I present here a mathematical model based on the influence of ACE2 on viral propagation and disease severity. The model fits Covid-19 fatality rate across age and sex with high accuracy ($r^2>0.9$) under the hypothesis that SARS-CoV-2 infections are in the dynamical regimes in which increased receptor slows down viral propagation. Moreover, rescaling the model parameters by the ratio of the binding rates of the spike proteins of SARS-CoV and SARS-CoV-2 allows predicting the fatality rate of SARS-CoV across age and sex, thus linking the molecular and epidemiological levels.
The presented model opposes the fear that angiotensin receptor blockers (ARB), suggested as a therapy against the most adverse effects of CoViD-19, may favour viral propagation, and suggests that Ang-II and ACE2 are candidate prognostic factors for detecting population that needs stronger protection.

\end{abstract}



The Covid-19 pandemics \cite{covid-19} is causing hundreds of thousands fatalities worldwide \cite{jhu}, creating a tremendous threat to global health. 
It presents a strong gradient of fatalities across age and a sex bias with much higher severity in males than females.
%
Modelling studies that extrapolate the number of infections suggest that, at young age, most SARS-CoV-2 infections are asymptomatic and the fatality ratio is very low, whereas for the elderlies most infections are severe and a large fraction of them can be fatal \cite{Symptoms_by_age}.
Understanding the biological reasons that underlie these striking differences is one of the most pressing problems of CoViD-19 research, which might lead to better prediction of the disease prognosis and possible treatments that approach the severity of the worst affected groups to that of the most protected ones.
 

Here I show that the case fatality ratio of Covid-19 across age and sex correlates negatively with the level of the protein Angiotensin converting enzyme 2 (ACE2), the cellular receptor both of SARS and SARS-CoV-2 virus \cite{covid-19,ACE2-SARS}. The correlation is very strong with membrane-bound ACE2 protein in rat lungs, which decreases with age and is higher in old females than old males \cite{ACE2-expression}.
The same qualitative pattern is observed for membrane-bound ACE2 in mice \cite{Mouse_ACE2}, where all the anti-inflammatory axis of the Renin-Angiotensin-System (RAS) \cite{RAS-review} to which ACE2 belongs decreases with age. In humans, ACE2 protein in serum \cite{ACE2-serum} and mRNA \cite{gtex-preprint,ACE2-children_nose, ACE2_expression_children,ACE2-mRNA-single-cell} are lower in children than adults, but available data and current knowledge of the RAS supports that ACE2 as membrane protein decreases through age \cite{ACE2_protein_children}, as hypothesized here and discussed later.

%

The negative correlation between ACE2 and lethality is surprising: higher levels of the receptor decrease the lethality exponentially. The paradox is only apparent because high receptor levels do not necessarily favour viral propagation.
A mathematical model of viral infection \cite{PRL_virus_propagation}, developed before the COVID-19 pandemics, computed how viral propagation depends on the adsorption rate of viruses on cells. In terms of receptor level, this model predicts that the speed of viral propagation is a non-monotonic function that increases with the receptor level, reaches a maximum and then decreases.

The second mechanism that may underlie the negative correlation concerns the function of ACE2 not as viral receptor but as key enzyme of the RAS \cite{RAS-review}. 
This system, besides regulating blood pressure (BP) and electrolyte homeostasis in blood, plays a central role in inflammatory processes \cite{ACE2_cytokine_1},
immune response \cite{ACE2_inflammation_loop} and coagulation \cite{hypertension-prothrombotic,ACE2-prothrombotic}, which characterize the most severe Covid-19 cases \cite{RAS_SCOV_axis,SARS_coagulopathy}. 
Its main player is the family of peptides derived from angiotensin I (Ang1-10), cleaved by the enzyme Renin from the protein angiotensinogen. Its pro-inflammatory arm is constituted by angiotensin II (Ang1-8), cleaved from Ang1-10 by the angiotensin converting enzyme (ACE) homologous to ACE2. Ang1-8 bound to the receptor ATR-1 triggers a cascade of reactions leading not only to vasoconstriction and increased BP but also to activation of the transcription factor NFkB that upregulates inflammatory cytokines (IL-1, TNF-$\alpha$ and IFN-$\gamma$ among others), activates white blood cells and platelets, and favours thrombotic processes \cite{NFkB}. The enzyme ACE2 limits the level of Ang1-8 by converting its precursor Ang1-10 to Ang1-9 \cite{ACE2-function} that is subsequently cleaved by ACE to Ang1-7, and by directly converting Ang1-8 to Ang1-7, which belongs to the anti-inflammatory arm of the RAS since it favours vasodilation, reduces BP and attenuates inflammation \cite{Ang1-7}.
 
%
Upon viral entry the spike protein of SARS-CoV and probably also SARS-CoV-2 cause the internalization and degradation of ACE2 \cite{Kuba} that critically contributes to lung damage \cite{Imai2005,ACE2-lung-1,ACE2-lung-2}.
Decrease of ACE2 raises the severity of lung injury in other inflammatory diseases \cite{ACE2-lung-2} and in aging rats \cite{Schouten2016}, which may be explained by the increase of Ang1-8 and its adverse effects.


Here I develop a set of mathematical models of SARS-CoV-2 lethality versus the pre-infection level of ACE2 based on two aspects: the influence of ACE2 on viral progression \cite{PRL_virus_propagation} and the negative effect of its degradation. These models are fitted to the CFR of SARS-CoV-2 across six classes of age and sex in Italy, Spain and Germany, and support the hypothesis that the receptor level slows down the virus propagation, which fits the data much better than the competing hypothesis.
%
Furthermore, under the same hypothesis and by rescaling the parameters fitted to SARS-CoV-2 by the ratio between the binding rates of the spike proteins of SARS-CoV and SARS-CoV-2, the model predicts well the CFR of SARS-CoV 
further supporting the negative relationship between receptor level and virus propagation.

\section*{Results}
\subsection*{CoViD-19 lethality correlates negatively with ACE2 level}

The level of the ACE2 protein in rat lungs were quantified across three age classes of the two sexes by Xie et al. \cite{ACE2-expression}, who found that it strongly decays with age and it is higher in female than male rats, with largest difference in the oldest cohort where the expression is almost double for females.
Here I assume that human ACE2 membrane protein levels across age and sex are qualitatively similar to rat data, apart for multiplicative factors that may depend on the organ.
This assumption is supported by mice data \cite{Mouse_ACE2}, ACE2 protein in human lungs \cite{ACE2_protein_children} and ACE2 mRNA expression in the GTEx database \cite{gtex-preprint}, but it apparently contrasts with the observation that ACE2 in serum \cite{ACE2-serum} and ACE2 mRNA \cite{ACE2-children_nose,ACE2_expression_children,ACE2-mRNA-single-cell} is higher in adults than in children. We discuss later that these discrepancies are in fact consistent with the current knowledge of the RAS.

\begin{figure}[!t]
\centerline{\includegraphics[width=0.8\linewidth]{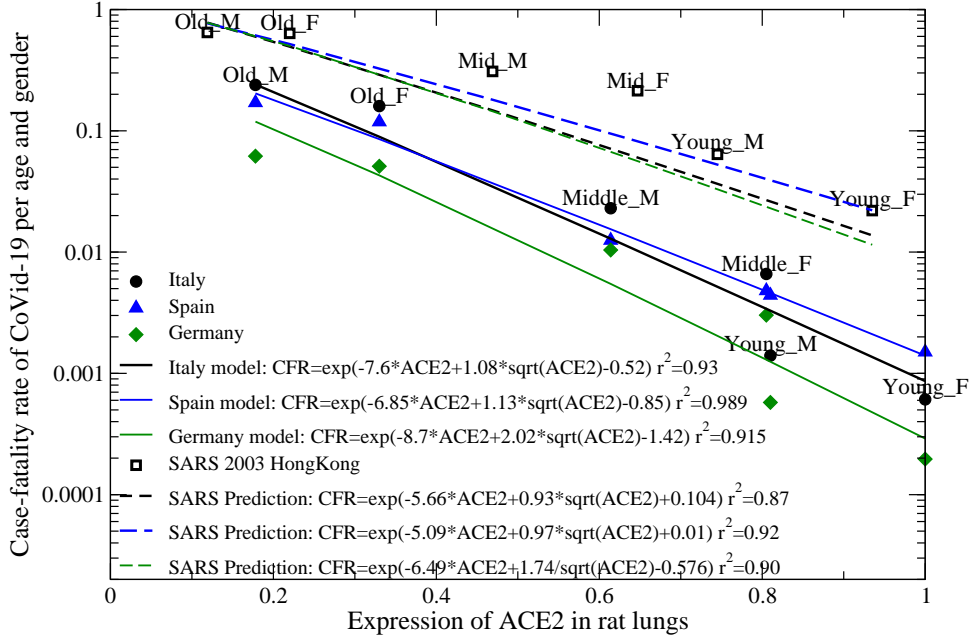}}
\caption{Expression of the ACE2 protein in lung rats (horizontal axis), normalized so that the highest expression is one, versus case fatality rate (vertical axis) of SARS-CoV-2 (Circles: Italy; triangles: Spain; diamonds: Germany) and SARS 2003 (open squares) in three age classes (young 0-29, middle-age 30-59 and old $>59$) and two sexs (male and female). The solid lines represent fits to the mathematical model (see text), the dashed lines represent predictions that rescale the models fitted to SARS-CoV-2 with the ratio between the binding rate constants of SARS and SARS-CoV-2 (see text).}
\label{fig:ace2}
\end{figure}

Strikingly, the profile of ACE2 in Ref.\cite{ACE2-expression} is very strongly anti-correlated with the lethality of SARS-CoV-2. Fig.\ref{fig:ace2} represents the level of the ACE2 protein in lung rats (horizontal axis, data from Fig.2 of \cite{ACE2-expression}) versus the case fatality ratio (CFR) of CoVid-19 registered in Italy \cite{ISS}, Spain \cite{CFR_Spain} and Germany \cite{CFR_Germany} in three uniform age classes (young, $<30$, middle-age $30-59$ and old $> 60$; vertical axis) of each sex. Data strongly support the exponential decrease of mortality with ACE2 level, with $r^2=0.91$, $0.97$ and $0.89$, respectively, suggesting that variations of ACE2 describe the largest part of the variation of the CFR.

As for other two-parameter fits tested in this work, the fitted exponents for Italy and Germany coincide within the error and the CFR differ only by a multiplicative factor, supporting the robustness of the data. Data from Spain present higher mortality in the young ages, which might be attributed to more frequent undetected cases in young age with lower severity.
%

\subsection*{Mathematical model of covid-19 lethality}

Mathematical models of viral growth consider three processes: virus adsorption into susceptible cells, production of virus by infected cells after a delay time $\tau$, and viral clearance by the immune system \cite{influenza_review}. I translate here this mathematical formalism in terms of receptor density, exploiting that the adsorption rate is proportional to the receptor level $A$ expressed in susceptible cells times the association rate between the virus and the receptor, $k_A\equiv k A$. 

The simple mean-field model that does not consider explicit space predicts a minimal receptor density below which the virus does not grow and above which higher receptor levels accelerate the viral progression. 
Considering spatial diffusion modifies this situation. The analytical solution of a model of viral propagation in space was obtained in 2002 by Fort and M\'endez \cite{virus_spatial_analytical}, who tested their model with experiments on the spread of bacteriophages in lysis plaques. Their mathematical formulation is conceptually equivalent to the present setting and can be directly applied here. Varying the receptor level $A$ (adsorption rate in the original paper), three regimes appear: (1) When $A$ is small the viral velocity $v$ increases with $A$ less than linearly. (2) For intermediate $k A$, large with respect to $1/\tau$ but small with respect to the rate of virus production, the viral velocity reaches a plateau where it is almost independent of $A$. (3) Although not explicitly discussed in Ref.\cite{virus_spatial_analytical}, the formulas presented there remain valid in the regime where $k A$ is larger than the rate of virus production. In this regime the viral progression slows down with receptor density as $v\propto1/\sqrt{k A}$.

The latter result is surprising: how can the virus progress more slowly for increasing receptor level? Since this is a mathematical model, the answer is readily found: in the model, viral particles are consumed when they enter a cell but the viral yield per infected cell does not increase when a cell is infected multiple times. It was in fact proposed that multiple viral entries in the same cell interfere with viral replication. Several viruses such as HIV \cite{HIV-receptor-downregulation,HIV_receptor}, measles \cite{measles_receptors}, influenza \cite{influenza_receptor} and hepatitis B \cite{hepB_receptor} downregulate their own receptor, preventing multiple entries. 
%
%
The mathematical result that, after the infection is established, very fast adsorption does not favour the virus, agrees with a recent study that demonstrated the protective effect of high adsorption rate through analytic computation, simulation and experiment \cite{phagus_PNAS}.
In all, the mathematical model predicts that viral progression in the organism declines above some receptor level.

Next, I consider two possible models of Covid-19 death. In the first one, death occurs when the virus propagates through the upper respiratory tract or through endothelial cells, reaches the lungs and infects and destroys a critical fraction $X$ of it, the same for all patients. The second model considers the protective effect of ACE2, assuming that the critical fraction $X$ is a decreasing function of the pre-infection ACE2 level $A$, $X=1-A_c/A$, i.e. the larger is the initial level of ACE2, the more tolerant is assumed to be the organism to the viral infection.

Combining these two assumptions with the three regimes of viral propagation described above gives six mathematical models. For each of them, I compute the time $t_d$ after which the virus causes death versus the ACE2 level $A$ (see Methods).
If the viral velocity increases with ACE2 $t_d$ is a decreasing function of $A$. This behaviour contradicts the data and I shall not consider it further. In the regime in which the viral velocity is independent of $A$ $t_d$ is an increasing function of $A$ only for the model that considers the protective effect of ACE2, which is called Model 1. Model 2 assumes that viral propagation decreases with $A$ but neglects the protective effect of ACE2 and Model 3 considers both the decrease of viral propagation and the protective effect of ACE2. 

I compute lethality as the probability that $t_d$ is smaller than the time $t_i$ needed by the immune system for clearing the virus, which I model as a random variable with two possible distributions: (1) The exponential distribution, which is the distribution with maximum entropy for given average value. (2) The Gaussian distribution, justified by the fact that $t_i$ is the sum of intermediate steps in the maturation of the immune response and the limit of the sum of independent random variables is a Gaussian. I adopt the same immune system parameters for all age and sex classes in order to limit the number of free parameters and to test if variations of the receptor level are sufficient to explain the lethality.

The three models, combined with the exponential and the Gaussian distribution, yield six different functional forms of the CFR versus $A$. Some of them have many parameters, but simple approximations reduce the number of free parameters to two (for the exponential distribution) or three (for the Gaussian distribution), see Methods.

The three models are fitted to the data under the same distribution, strongly regularizing the fits with rescaled ridge regression \cite{RRR} to limit overfitting as much as possible (see Methods).
Under both distributions, and for the CFR of the three countries, model 1 gives the worst fit and model 3 the best. In particular, under the exponential distribution the relative quadratic error of the logarithm for models 1, 2 and 3 is $0.37$, $0.19$, $0.07$ for Spain, $0.41$, $0.30$, $0.21$ for Italy and $0.48$, $0.23$, $0.17$ for Germany, using two free parameters in all cases. Adopting the Gaussian distribution requires three parameters, with high risk of overfitting, but it maintains the same ranking: $0.04$, $0.01$ and $0.01$ for Spain, $0.10$, $0.04$ and $0.04$ for Italy and $0.14$, $0.06$, $0.06$ for Germany, see Methods and Supplementary Table S1. 

In conclusion, for both distributions the hypothesis that the viral progression decreases with the receptor level (models 2 and 3) fits the data much better than the competing hypothesis that the propagation is independent of the receptor level. Under the exponential distribution model 3, which also considers the protective role of ACE2, performs better than model 2, but they are undistinguishable under the Gaussian distribution.

The solid lines in Fig.\ref{fig:ace2} represent approximate fits of model 2 under the Gaussian distribution: $-\log(\mathrm{CFR})=aA-b\sqrt{A}+c$. They yield $r^2>0.92$ with only two free parameters, since the parameter $c$ was not fitted to reduce the error of the fit (see Methods). The parameters $a$ and $b$ fitted from different countries differ only slightly more than their statistical error, which may reflect the different incidence of undetected cases across age classes.


\subsection*{SARS 2003}

An important prediction of Models 2 and 3 is that the fit parameters $a$ and $b$ depend on the adsorption constant per unit receptor $k$, with the fit parameters expressed as $a\propto k$ and $b\propto \sqrt{k}$ in Model 2. Crucially, this prediction can be tested on the CFR of the 2003 SARS coronavirus \cite{SARS-HongKong-2}, which also uses ACE2 as cellular receptor. Under the hypothesis that the rate-limiting step for adsorption is the binding of the spike protein to ACE2, justified if the complex is stable enough to allow membrane fusion, and supported by infection assays, $k$ is proportional to the spike binding rate constant, which has been measured with biochemical experiments for both SARS and SARS-CoV-2. This allows rescaling the fit parameters $a$ and $b$ obtained for SARS-CoV-2 to predict the corresponding parameters for SARS (see Methods).
%
Multiplying the predicted CFR times a global factor that accounts for undetected cases as the only free parameter, we predict the CFR of SARS with very good accuracy. The relative quadratic error is equal to $0.13$, $0.08$ and $0.01$ using the fit parameters from Italy, Spain and Germany, respectively, see Fig.\ref{fig:ace2}, Supplementary table S1 and Discussion. This result further supports the hypothesis that the viral velocity is slowed down by increasing receptor level.

\subsection*{NL63}

It is natural to extend this analysis to the other human coronavirus that uses ACE2 as receptor, NL63 that causes common cold and is not generally associated with pneumonia. Its spike protein contains a very stable receptor binding domain of 120 residues that showed high binding affinity for ACE2 \cite{NL63_spike_sructure}. However, the complete S1 domain of the spike (717 residues) is much less stable and its affinity for ACE2 is so small that it could not be measured with binding assays \cite{NL63,NL63_ACE2_affinity,NL63_ACE2_downregulation}, and it was conjectured that it is 10-100-fold smaller than that of SARS-CoV \cite{NL63_ACE2_affinity}.
Since the CFR of SARS peaks for old females, whose normalized ACE2 is equal to 22\% of the maximum value, the model predicts that this is the level at which SARS-CoV propagates fastest. If the binding affinity of the NL63 spike is at least ten times smaller, the ACE2 level at which NL63 propagates fastest is more than double the highest ACE2 level, implying that NL63 is in the regime in which the ACE2 level enhances its propagation.

This analysis agrees with the apparently surprising data reported in Fig.3A of Ref.\cite{NL63}, which shows that ACE2 overexpression in 293T cells enhances NL63 infection three times more than SARS-CoV infection, indicating that higher receptor levels accelerate the propagation of NL63 more than that of SARS-CoV despite the latter has higher binding affinity.
%


\section*{Discussion}

Since ACE2 is the SARS-CoV-2 receptor, we may expect that raising its level enhances the rate at which the virus propagates in the organism and worsens the outcome of the infection.
However, a mathematical model of viral progression presents a regime in which the increase of the receptor level slows down the virus propagation in the organism. The observed relation between SARS-CoV-2 lethality and ACE2 levels suggests that this is the relevant regime of SARS-CoV-2 infections, as further supported by the prediction of the age- and sex-dependent lethality of 2003 SARS-CoV.
%


\subsection*{Human ACE2}

An important limitation of the present work is that it uses data of ACE2 protein levels in healthy rat lungs \cite{ACE2-expression} since similar data are not available for humans. A recent clinical study could not confirm significant differences of ACE2 expression between patients of different age with acute respiratory distress syndrome (ARDS) \cite{Age-dependent_ACE2_clinic}. This clinical study measured the activity of RAS enzymes at only one time point for each patient, so that the dynamics of the RAS during ARDS may have obscured individual differences.

Consistent with rat data, decrease through age of membrane-bound ACE2 protein was also observed in mice, which show a general strengthening of the inflammatory arm of the RAS with aging \cite{Mouse_ACE2}. In humans, a recent preprint found that membrane-bound ACE2 is more abundant in children than in adult lungs \cite{ACE2_protein_children}.
However, other publications and preprints showed that ACE2 is higher in adults than in children in serum \cite{ACE2-serum} and at the mRNA level \cite{ACE2-children_nose,ACE2_expression_children,ACE2-mRNA-single-cell}.
ACE2 is detached from the cellular membrane and shed to the serum by the metalloprotease ADAM17 \cite{ADAM17}.
The discrepancy between membrane-bound and serum levels suggests that ADAM17 is less active in children, consistent with our current and still limited knowledge of the RAS. In fact, ADAM17 is upregulated by the binding of Ang1-8 to the angiotensin II type 1 receptor (ATR1). Ang1-8 increases with age \cite{Mouse_ACE2}, while children express much more than adults the receptor ATR2 that competes with ATR1 and counteracts its action \cite{ATR2_2017}. Thus, children are expected to present lower binding of Ang1-8 to ATR1 and lower activation of ADAM17. Regarding ACE2 mRNA, this starts being expressed after birth and reaches a maximum at young age \cite{ACE2-mRNA-single-cell}. The hypothesized lower shedding of ACE2 in children implies that the membrane protein reaches its maximum at younger age than mRNA. Therefore, the higher expression of ACE2 as membrane protein is not contradicted by these observations.

Rat data are also consistent with the observation that ACE2 mRNA expression decreases through age in mice \cite{Mouse_ACE2_2} and adult human cells \cite{gtex-preprint}, as also represented in Fig.3g of the preprint that reported higher ACE2 mRNA in adults \cite{ACE2-mRNA-single-cell}. Mechanistically, this decrease of ACE2 mRNA through age is probably related with the increase of Ang1-8 \cite{Mouse_ACE2} that downregulates ACE2 expression \cite{Angiotensin_feedback}. 

Regarding sex differences, Ref.\cite{gtex-preprint} found that ACE2 mRNA is higher in males than females, as in rats. Ref.\cite{ACE2-mRNA-single-cell} reached the opposite conclusion, but this seems an artefact of the fact that smoking enhances ACE2 expression \cite{ACE2_expression_children} and in their samples 50\% of men were smokers compared to 25\% of women.
%
It has to be noted that the ACE2 gene is contained in the X chromosome, of which females have two copies. Although one of these copies is epigenetically silenced, about 15\% of the X-linked genes escape this inactivation \cite{X-inact} and heterochromatin is known to dysregulate with age. It is interesting that old female rats present almost exactly double ACE2 than males \cite{ACE2-expression}, as one would expect if the epigenetic silencing fades at old age.
Consistently, some of the sex differences in human cardio-vascular diseases have been attributed to sex differences in the expression of ACE2, which acts as protective factor \cite{ACE2-cardio}.

The negative relation between ACE2 levels and severity of Covid-19 is supported by other risk and protective factors corrected for age, sex and other comorbidities in a large study in the UK \cite{comorbidities}. Namely, being a current smoker constitute a curious protective factor (adjusted hazard ratio (AHR): $0.82-0.97$), and smoking enhances ACE2 expression \cite{ACE2_expression_children}. Contrary to single-factor analysis, diagnosed hypertension is a protective factor (AHR $0.85-0.93$), which may be correlated with the fact that anti-hypertensive drugs enhance ACE2 expression \cite{ARB-Ferrario}. Diabetes is a risk factor (AHR $1.72-2.09$), and it has been associated with reduced ACE2 expression \cite{Diabetes-RAS}. Cardiovascular diseases and reduced kidney function are additional risk factors that are related with reduced ACE2 levels \cite{RAS-review}. Finally, Vitamin D deficiency is a risk factor for COVID-19 \cite{VitaminD} that is also related with low levels of ACE2 because Vitamin D inhibits the expression of Renin, which in turn produces Ang1-10, the substrate from which is cleaved Ang1-8, which downregulates ACE2 \cite{VitaminD-RAS}. Therefore, low levels of Vitamin D are expected to reduce ACE2. Vitamin D levels are decreased in ethnic groups with dark skin pigmentation living at temperate latitudes, perhaps due to high screening of solar radiation, providing a possible causal relationship between ethnic status and Covid-19 (AHR $1.30-1.69$ for Black people, discounting socio-economic factors), once again through ACE2. Therefore, other hazard factors and protecting factors besides age and sex also support a negative correlation between ACE2 and Covid-19 lethality.

Finally, the GTEx database shows that, despite lungs are the organ that is more severely damaged by COVID-19, they do not present high expression of ACE2 mRNA \cite{gtex}, which is higher in tissues from reproductive organs, intestine, adipose tissue, kidney, hearth, thyroid, esophagus, breast, salivary glands and pancreas, among others. Some of these organs may be infected but they experience less damage, consistent with the negative correlation between ACE2 levels in lungs and lethality.

\subsection*{Role of ACE2 for virus propagation and spike mutant D614G}

The above evidence strongly supports the negative correlation between ACE2 protein levels and severity of CoViD-19. This in turn supports the mathematical model presented here, based on the hypothesis that increased ACE2 slows down viral propagation \cite{virus_spatial_analytical}, which fits the CFR from Spain, Italy and Germany with $r^2=0.93$, $0.79$ and $0.83$, respectively, using two free parameters.
The same hypothesis predicts the lethality profile of the 2003 SARS virus across age and sex yielding $r^2=0.92$, $0.87$ and $0.99$ using the fit parameters from Spain, Italy and Germany, respectively, and a single free parameter.
This extrapolation from SARS-CoV-2 to SARS uses the ratio between the binding rate constants of the spike proteins of the two viral species, bridging the molecular and the epidemiological level.

The above computation predicts that SARS has higher relative mortality for young age with respect to old age (the observed ratio is 22\% for SARS compared with 1.3\% for SARS-CoV-2) due to the smaller binding affinity of the SARS spike for ACE2. It also predicts that mutations that decrease the binding of SARS-CoV-2 spike may generate a strain more severe for younger age.

Remarkably, the mutant D614G of the viral spike, which rapidly rose to almost fixation world-wide \cite{D641G_fixation}, presents lower affinity for ACE2, propagates faster in cell cultures than the original spike \cite{D641G_conformation} and its detected cases tend to be younger \cite{D641G_lethality}, in line with the above prediction. A direct relation between D614G and disease severity could not be proven, but an indirect one seems to exist since D641G is associated with increased viral load and viral load is associated with hospitalization \cite{D641G_lethality}.
Nevertheless, the improved propagation of D614G was attribute to the higher population of the binding-competent open configuration \cite{D641G_conformation}, thus other possible interpretations exist.



\subsection*{ARB and ACE2 as protective factors}

SARS and probably also SARS-CoV-2 degrade ACE2 \cite{Kuba}, with detrimental effects on the lungs on which ACE2 has a protective effect \cite{Imai2005,ACE2-SARS-rev,ACE2-lung-1,ACE2-lung-2}.
Several papers proposed that the downregulation of ACE2 is a key factor for the severity of CoViD-19 and suggested that ACE inhibitors (ACE-I) and angiotensin receptor blockers (ARB) that limit the effects of Ang1-8 may be beneficial for CoViD-19 patients \cite{ARB_proposal,Gurwitz,ACE2-new-1,ACE2-new-2,ARB_Offringa,ARB_editorial}. A similar idea was already proposed at the time of SARS, and a retrospective meta-analysis found that the use of ARB and ACE-I provides a consistent reduction in risk of pneumonia compared with controls 
\cite{ACEI-inhib-review}. In the context of CoViD-19, several studies found that ARB and ACE-I protect the lungs and mitigate the severity of COVID-19 for patients that already take them against hypertension \cite{ARB-clinic-1,ARB-clinic-2,ARB_clinic_Italy,ACE2-SARS2-debate,RAS-COVID19-debate}.

%
%

However, the possible protective role of ARB and ACE-I has been out-weighted by the fear that they may favour viral propagation since they upregulate the viral receptor ACE2 \cite{ARB-Ferrario}, and it was proposed that they should be discontinued \cite{Fang,Adverse_ARB}. Medical societies firmly opposed this suggestion due to lack of evidence \cite{ESC,EMA,HFSA}.
%
%
The current consensus is that available data are too limited to support either hypothesis that ACE-I and ARB may be detrimental or beneficial in CoViD-19 infections, but withdrawal of anti-hypertensive drugs in patients that need them may be harmful \cite{ACE2-SARS2-debate}.
%

The negative correlation between ACE2 levels and lethality of SARS-Cov-2 found here, and the mathematical prediction that the receptor level does not enhance but slows down viral progression, contradict the fear that ARB and ACE-I may benefit the virus and suggests two complementary protective roles of high ACE2 levels. On one hand, they may slow down the propagation of the virus, an effect conceptually similar to that observed in recent experiments with soluble human ACE2 \cite{ACE2-soluble}. On the other hand, they reduce the accumulation of Ang1-8, whose proinflammatory and prothrombotic effects are thought to underlie the most severe complications of CoViD-19.
This work thus supports the idea that ARB and ACE-I used to treat high blood pressure may limit the most adverse manifestations of CoViD-19.

A note of caution concerns the effect of these drugs on the bradykinin system. This system is strongly coupled with the RAS and causes vasodilation, reduces blood pressure and increases vascular permeability. Its overactivity can lead to increased inflammation, thrombosis and angioedema in the lungs, and it was proposed that it mediates the severe manifestations of COVID-19 \cite{Bradykinin1,Bradykinin2,Bradykinin3}.

The bradykinin system consists of two axes. The first one is downregulated by ACE2, which degrades the signalling peptide des-Arg$^9$-bradykinin (DABK) whose receptor BK1R is upregulated by Ang1-8 (in turn downregulated by ACE2) bound to the receptor ATR1. Stimulation of this axis may lead to release of pro-inflammatory chemokines, lung inflammation and injury \cite{B1Raxis} and is expected to be reduced through ARB and ACE-I, which would exert a protective role.

The other axis is downregulated by ACE, which degrades the signalling peptide BK, whose receptor BK2R is in turn activated by Ang1-7 and Ang1-9 produced by ACE2, and is stimulated by Ang1-8 bound to the receptor ATR2 \cite{ATR2-Bradykinin}. Thus, ACE2, ATR1 receptor blockers and even more ACE-I can upregulate the BK2R axis with pathological consequences, as observed in severe side-effects of ACE-I \cite{Bradykinin-ACEI}, and their use should be limited in the presence of hypotension. Nevertheless, adverse effects have not been reported in studies of the effect of these drugs on COVID-19 patients. We speculate that the synergy between BK2R and ATR1 through heterodimerization \cite{BK2R-AT1R} might generate a negative feedback that limits the effect of ARB on the BK2R axis.

\subsection*{Positive feedback loop}
It is noteworthy that degradation of ACE2 increases the level of Ang1-8, which in turn binds the ATR1 receptor and down-regulates ACE2 even further both at the mRNA and at the protein level \cite{Angiotensin_feedback}.
Thus, the SARS-CoV-2 infection may trigger a dangerous positive feedback loop that strongly raises Ang1-8, exacerbating inflammatory response \cite{ACE2_cytokine_1,ACE2_inflammation_loop} and coagulation problems \cite{hypertension-prothrombotic,ACE2-prothrombotic}, frequent complications of severe CoViD-19 patients
\cite{SARS_coagulopathy,Covid-inflammation}.
Positive correlation between Ang1-8 levels and viral load has been reported in CoViD-19 patients \cite{Angiotensin_viral_load}, supporting the link between severe CoViD-19 and dysregulation of the RAS.

Under this point of view, ARBs appear to be more favourable than ACE-I because they can interfere with the positive feedback loop of Ang1-8 and because Ang1-8 can be generated by other proteases if ACE is inhibited \cite{RAS-review}.

\subsection*{Concluding remarks}
Of course, clinical trials are necessary to establish whether ARB and ACE-I have a positive, negative or neutral effect on CoViD-19 severity. To this aim, the clinical trials NCT04312009 and NCT04311177 started in April 2020 at the University of Minnesota. The present work aims at stimulating others to join efforts to assess this promising treatment.


Finally, the results presented here suggest a prognostic role for the measurements of ACE2 levels in bronchial aspirated lavage samples and Ang1-8 in the serum, which may predict the severity of the disease already at an early stage and may allow detecting risk groups that need higher protection besides the elder, as supported by the association between ACE2 and known risk and protecting factors against CoViD-19 \cite{comorbidities}. We are currently investigating this possibility through retrospective studies.

\section*{Materials and Methods}


\subsection*{Case-fatality-rates and expression data}
Case fatality rates (CFR) were taken from Ref.\cite{ISS,CFR_Spain,CFR_Germany} for CoViD-19 in Italy, Spain and Germany, respectively, and from Ref.\cite{SARS-HongKong-2} for the 2003 SARS outbreak in Hong-Kong. At the beginning of an outbreak, CFR underestimate the true fatality rate because their calculation assumes that all people currently infected will recover, which unfortunately is not true. This effect may not be uniform across age-sex classes if patients of some classes tend to die more rapidly, as assumed by the model. However, at a late epidemic stage this effect is expected to be small. On the other hand, CFR overestimate the true fatality rate because of undetected cases that tend to lower the denominator. Since age-sex classes with higher lethality also tend to have more severe cases and less undetected cases, the overestimation is larger for classes with smaller lethality, with the consequence of reducing the differences among classes for larger fraction of undetected cases. The data currently available do not allow correcting for this bias, which may account for some of the differences in the fit parameters.

Expression data presented in Ref.\cite{ACE2-expression} were grouped in three age classes of 3 (young), 12 (middle) and 24 months (old). CFR were presented in bins of 10 years, and I grouped them in three equally spaced groups 0-29 (young), 30-59 (middle) and $\geq 60$ years (old). Grouping the 20-29 years class with the middle age gave similar results with approximately exponential decrease of lethality with ACE2 expression.

For SARS CFR \cite{SARS-HongKong-2}, ages were grouped differently: 0-44 (young), 45-74 (middle) and $\geq 75$ (old). To compare these groups with those of SARS-CoV-2, I interpolated expression data of ACE2 $A$ for these groups as $A(0-44)=0.667 A(0-29)+0.333 A(30-59)$, $A(45-74)=0.5 A(30-59)+0.5 A(\geq 60)$ and $A(\geq 75)=0.667 A(\geq 60)$. Other schemes gave qualitatively similar results: The CFR decreases approximately exponentially with $A$ and the exponent is smaller than for SARS-CoV-2, which are the two important points made in the paper.

\subsection*{Mathematical model of viral propagation}
The simplest mathematical model of viral propagation considers three populations: uninfected cells $U(t)$, free virus $V(t)$ that enter the cells with rate $k_A U(t)V(t)$ (adsorption) and are cleared with rate $c$, and infected cells $I(t)$ that produce new virus at rate $k_V Y I(t)$ ($Y$ stands for yield) after a delay $\tau$ called eclipse time, until they ultimately die \cite{influenza_review}. Here I express the adsorption rate as a function of the receptor density $A$, $k_A=k A$.
The viral population cannot grow below the minimum receptor density given by $A^\mathrm{min}=c/(k U_0 Y)$ ($U_0$ is the initial concentration of susceptible cells).
A spatially explicit version of this model in which virus diffuse through susceptible cells \cite{PRL_virus_propagation} was found amenable to analytical solution that explicitly gives the viral velocity $v$ as a function of the model parameters \cite{virus_spatial_analytical}. Here I describe this solution in terms of the receptor density. The authors describe two regimes: (1) For small $A$ ($A<1/(k U_0 \tau Y)$, $A<1/(k U_0 k_V)$), the viral velocity increases with $A$, but less then linearly, as $v=2\sqrt{D\frac{k_A U_0 Y}{1+k_A U_0 Y}}$.
(2) For intermediate receptor density $1/(k U_0\tau Y)<A<1/(k U_0 k_V)$ the viral velocity is given by $v=\sqrt{2D/\tau}$, where $D$ is the effective diffusion constant that depends on cell shape and density, and it is almost independent of $A$ \cite{virus_spatial_analytical} so that the viral progression is not enhanced by the expression of the receptor.
(3) The formulas presented in the paper are also valid in the third regime of very high receptor density, when $k A$ is larger than the rate at which viral particles are produced: $A>1/(k U_0\tau Y)$, $A>1/(k U_0 k_V)$, although this regime was not explicitly discussed. Counter-intuitively, in this regime the viral progression slows down with receptor density as $v=\sqrt{k_V/k A U_0}$.

The time that it takes for the virus to propagate through the upper respiratory tract (URT) can be estimated as $t_U=l_U/v$.
When the virus reaches the lungs, the number of infected cells grows with time as $I(t)\propto (vt)^{d_F}$, where $d_F\approx 2.35$ is the fractal dimension of the lungs \cite{lung_fractal_Weibel},
which are one of the classical examples of a fractal organ. 
As cells get infected, the receptor density in the lungs decreases as $A(t)=A(0)\left(1-I(t)/l_L^{d_F}\right)$ and it reaches the critical level $A_c$ after the time $t_L=(l_L/v) (1-A_c/A_0)^{1/d_F}\approx =\frac{l_L}{v}\left(1-\frac{A_c}{d_F}\frac{1}{A(0)}\right)$. I used the approximation $A_c\ll A(0)$, and $A(0)$ is the receptor density at the beginning of the infection, which in the main text is simply denoted as $A$. Summing these two times, I estimate the time at which death occurs as $t_d \approx (1/v)\left[l_U+l_L\left(1-\frac{A_c}{d_F}\frac{1}{A}\right)\right]$.

I consider three situations: (1) $v$ is independent of $A$; (2) $v$ decreases as $1/\sqrt{k A}$ and almost all the cells in the lungs must be infected to produce the death, i.e. $A_c=0$; (3) $v$ decreases with $A$ and $A_c>0$. In each situation, the death time $t_d$ depends on $A$ as
$t_d\propto 1-\frac{A_c}{d_F}\frac{1}{A}$
(Model 1),
$t_d\propto \sqrt{k}\sqrt{A}$
(Model 2),
$t_d\propto \sqrt{k}\left(\sqrt{A}-\frac{A_c}{d_F}\frac{1}{\sqrt{A}}\right)$
(Model 3).

In the model, death occurs if $t_d$ is smaller than the time $t_i$ needed by the immune system to control the virus, which is modelled either as an exponential (E) or a Gaussian (G) random variable. In the first case, the probability that $t_i$ is larger than $t_d$ can be computed as $P_d=\exp(-t_d/T)$, where $T$ is the average value of $t_i$. In the Gaussian case, the probability is well approximated as $P_d=C\exp\left(-(t_d-\mu)^2/(2\sigma^2)\right)$. Combining these expressions with the three models of $t_d$ versus $A$ and grouping together terms with the same power of $A$, we obtain six mathematical models of the lethality $P_d$ as a function of the initial level of ACE2 $A$:

\begin{equation}
-\ln(P_d)\approx
\left\{
\begin{array}{ll}
-\frac{a}{A}+b\;\;\;\mathrm{(1E)} &
\frac{a}{A^2}-\frac{b}{A}+c\;\;\;\mathrm{(1G)} \\
a\sqrt{A}+b\;\;\;\mathrm{(2E)} &
aA-b\sqrt{A}+c\;\;\;\mathrm{(2G)} \\
a\sqrt{A}-\frac{b}{\sqrt{A}} \;\;\;\mathrm{(3E)} &
aA-b\sqrt{A}+\frac{c}{\sqrt{A}}\;\;\;\mathrm{(3G)} \\
\end{array}
\right.
\nonumber
\end{equation}
$a$, $b$ and $c$ are positive fitting parameters. In Eq.(3G), there are five terms proportional to $A$, $\sqrt{A}$, $1/\sqrt{A}$, $1/A$ and constant, corresponding to five fitting parameters. In order to fit only three parameters, the same number as for the other Gaussian fits, I neglected the terms proportional to $1/A$ and constant, obtaining Eq.(3G), while neglecting the terms proportional to $1/A$ and $1/\sqrt{A}$ yields Eq.(2G).

\subsection*{Fits of the model}
The fitting parameters $a$ and $b$ are determined through regularized fits performed with rescaled ridge regression \cite{RRR}, which minimizes the quadratic error plus the term $\Lambda(a^2+b^2)$ that penalizes large values of the parameters. The regularization is necessary to avoid amplifying the noise due to covariant explanatory variables, as in the present case, and it allows more robust parameter estimation, often avoiding that they acquire unphysical values with incorrect sign, at the price of some bias. Rescaled ridge regression yields non-vanishing parameters even in the limit of large $\Lambda$, overcoming a drawback of other regularization schemes, and it fixes the parameter $\Lambda$ based on an analogy with statistical mechanics at the transition between the phase dominated by the noise and the one dominated by the bias.

For ridge regression there is no analytic formula to determine the statistical error of the fitting parameters, therefore I applied a bootstrap approach, repeating the fit while eliminating each of the data points and computing the standard deviation of the fitting parameters numerically.

Models (1G,2G,3G) coupled to the Gaussian distribution depend on three parameters but only the parameters $a$ and $b$ were fitted while $c$ was fixed, fitting $-\ln(P_d)-c=aA-b\sqrt{A}$. The parameter $c$ was determined not by minimizing the fit error but by selecting the value of $c$ that yields 50\% relative error on the parameters $a$ and $b$. The plots shown in Fig.1 were obtained in this way.

\subsection*{Prediction for SARS}

The models fitted to SARS-CoV-2 were rescaled in order to apply them to SARS-CoV, adopting the ratio between the binding rates of the spike proteins of both viruses to ACE2.
The most precise measures available in the literature are $k_\mathrm{SARS-2}=(2.3\pm 1.4) 10^5 nM^{-1}s^{-1}$ and $k_\mathrm{SARS}=(1.7\pm 0.7) 10^5 nM^{-1}s^{-1}$ (table 1 in \cite{spike-cell}). Although the error bars are huge, the greater rate constant of SARS-CoV-2 agrees with the more precisely determined binding affinity from the same table ($K_\mathrm{SARS-2}=(1.2\pm 0.1) nM$ and $K_\mathrm{SARS}=(5.0\pm 0.1) nM$), and from Ref.\cite{spike} that indicates that the spike protein of SARS-CoV-2 has greater affinity for ACE than the one of SARS-CoV. In that paper only one experiment was performed instead of five in Ref. \cite{spike-cell} and the binding rate constant was greater for SARS-CoV, which is consistent with the large statistical errors measured in Ref. \cite{spike-cell}. Thus, although the available data is quite noisy, the best available evidence suggests that the binding rate of SARS-CoV-2 spike protein is on the average faster than for SARS and the binding is more stable, which may also contribute to faster adsorption giving the virus more time to perform membrane fusion.

For predicting the CFR of the 2003 SARS outbreak, I used the parameters of SARS-CoV-2 and rescaled them with the ratio between the kinetic binding constant $k^\mathrm{on}$ of the two spike proteins: $a_\mathrm{SARS}=a_\mathrm{SARS-2}/1.35$ and $b_\mathrm{SARS}=b_\mathrm{SARS-2}/\sqrt{1.35}$. I obtained the lethality profile as $\exp(-a_\mathrm{SARS}A+ b_\mathrm{SARS}\sqrt{A})$, where $A$ is the ACE2 level of each sex and age class, and multiplied it times a constant factor that accounts for the different fraction of undetected cases, which was the only free parameter determined through a fit.

\section*{Acknowledgements}
This work is supported by the Spanish Research Council (CSIC) under the grant 202020E165.
I would like to thank medical doctors, nurses, police and other public employees that fight against the virus at the risk of their lives. This work is dedicated to my father, who passed away few days before the coronavirus outbreak became manifested in Italy, and my mother, who suffered cardiovascular diseases. I acknowledge useful discussions with several colleagues, in particular Patrick Chambers, Esther Lazaro, Manuel Fresno, Alberto Pascual-Garcia, David Abia, Alberto Rastrojo, Mario Mencia, Laura Garcia-Bermejo and Fatima Sanchez.

\end{document}